# On the Achievable Rate Regions for a Class of Cognitive Radio Channels:

Interference Channel with Degraded Message Sets with Unidirectional Destination Cooperation

Hsuan-Yi Chu
Department of Electrical Engineering
National Taiwan University
b95901210@ntu.edu.tw

*Abstract*—This paper considers the capacity gains due to unidirectional destination cooperation in cognitive radio channels. We propose a novel channel, interference channel with degraded message sets with unidirectional destination cooperation (IC-DMS-UDC), to allow the receiver of cognitive radio (secondary user) to participate in relaying the information for primary system (legitimate user). Our main result is the development of an achievable rate region which combines Gel'fand–Pinsker coding with partial-decode-and-forward strategy employed in the relay channel. A numerical evaluation of the region in the Gaussian case is also provided to demonstrate the improvements.

## I. Introduction

It has been a trend to allow cooperation among multiple nodes or users to improve efficiency of spectrum utilization and increase transmission rates. Upon the existing literatures, the interference channel with degraded message sets (IC-DMS) has attracted great attentions recently. The IC-DMS refers to a communication model, where two senders attempt to communicate with their respective receivers through the common medium simultaneously, and one of the senders has full knowledge of the message of the other sender. The studies on IC-DMS are motivated by two communication scenarios, cognitive radio and sensor network. The former is a two-user communication system where one of the senders is noncausally given the other sender's message [9]. And the latter consists of spatially distributed sensors in a correlated field where each sensor possesses a nested message structure to cooperate with one another to transmit information. The central aspect of this channel lies in the fact that there is a kind of cooperation between two senders. Since only one of the senders knows the message of the other sender, some literatures (e.g., [8]) call this kind of cooperation as unidirectional cooperation. It was shown in [8]-[12] that with this cooperation, the achievable rate regions can be significantly expanded.

From an information-theoretic perspective, the IC-DMS has been investigated in [8]-[12]. Although several achievable rate regions have been established, the capacity region of this channel in general is still unknown except for two special cases characterized in [8], [10], [12]. Notably, in the coding scheme of [11], sender 2 (as shown in Fig. 2 in [11]) splits its message into two parts, public and private messages. While the private message should be decoded by the intended receiver only, the public message should be decoded by both receivers. The public message enables receiver 1 to decode part of the information sent from sender 2. Therefore, the effective interference at receiver 1 can be reduced. In addition, because sender 2 has full knowledge of the message of sender 1, it applies the well-known Gel'fand-Pinsker coding [5] to separately encode its two split messages by treating the codewords of sender 1 as known interferences. Remarkably, it has been proved in [11, Corollary 1] that the achievable rate region derived by this coding scheme includes the capacity region for the Gaussian IC-DMS in the *low-interference-gain* regime, where the normalized link gain between sender 2 and receiver 1 is less than or equal to 1.

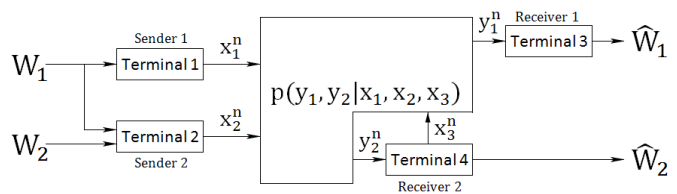

Fig. 1. Interference channel with degraded message sets with unidirectional destination cooperation (IC-DMS-UDC).

In this paper, we propose a novel channel, interference channel with degraded message sets with unidirectional destination cooperation (IC-DMS-UDC), depicted in Fig. 1. This channel is similar to IC-DMS. The distinction lies in the fact that receiver 2 is allowed to participate in relaying the information for sender 1 to enhance the cooperation between two users. We are motivated by two significant issues. First, it has been shown that cooperative communications can provide enormous benefits. Although there have been a lot of investigations on cooperation at senders, cooperation at receivers has attracted less attention. Second, in network, there are various ways and types of relays which can help communication. Nevertheless, less attention has been received on integrating the relaying strategies into IC-DMS. In our model, receiver 2 not only decodes the message sent from its sending dual but also relays the message sent from sender 1 to receiver 1. Although it may inevitably increase the design complexity of receiver 2, ultimately the achievable rate regions can be expanded, which results in more efficient utilization of frequencies.

There are various strategies for relaying information. For discrete memoryless channels with a single sender-receiver pair, three basic strategies were developed in [4]: *decode-and-forward*, *partial-decode-and-forward* and *compress-and-forward*. In our coding scheme, we employ partial-decoding-and-forward strategy to demand receiver 2 to decode part of the message sent from sender 1. In addition, sender 2 splits its message into public and private sub-messages and encodes both sub-messages by Gel'fand-Pinsker coding. With this coding scheme, we obtain our achievable rate region for the IC-DMS-UDC. We compare this result with the existing literatures and obtain a crucial conclusion: Given a constant power for transmitting the message of sender 1, if a part of this power can be allocated to receiver 2, which works as the relay for sender 1, the achievable rate regions can be expanded significantly over the case where all the power is centered at sender 1.

We note that V. Prabhakaran et al. studied interference channels with source cooperation in [13] and with destination cooperation in [14]. In addition, a subtle relation, called reversibility, between these two settings is discussed in [13, Section 7.1]. In this paper, we focus on unidirectional destination cooperation in IC-DMS, which is fundamentally different from the interference channels studied in [13], [14].

This paper is structured as follows: In Section II, the notational conventions and the channel model of IC-DMS-UDC are defined. In Section III, we present an achievable rate region for IC-DMS-UDC which is the main result of this paper. In Section IV, we discuss two special cases of this achievable rate region and a numerical example is provided to demonstrate the improvements. In Section V, we conclude the contributions of this paper.

## II. NOTATIONS AND CHANNEL MODEL

### A. Notations

The following notational conventions will be adopted throughout this paper. We use capital letters to denote the random variables and lower case letters to denote their corresponding realizations. And we adopt the notational convenience $P_{Y|X}(y|x) = P(y|x)$ to drop the subscript of probability distribution. We write $X \sim P(x)$ to indicate that $X$ is drawn according to $P(x)$. Moreover, we use $X^n$ to denote the random vector $(X_1, X_2, X_3, \ldots, X_n)$. And we use $|\cdot|$ to denote the cardinality of a set. For the information theoretic quantities such as entropy, mutual information, typical sequences, etc, we follow the notations of [2].

### B. Channel Model

The IC-DMS-UDC, depicted in Fig. 1, is a channel, where sender 1 sends a message $w_1 \in W_1 = \{1,2,3,\ldots,|W_1|\}$ to its receiver with the help of receiver 2 as a relay in n channel transmissions and sender 2, with full knowledge of the message of sender 1, sends a message $w_2 \in W_2 = \{1,2,3,\ldots,|W_2|\}$ to its receiver in n channel transmissions. Additionally, we focus on discrete memoryless channel. The channel is said to be discrete memoryless in the sense that

$$p(y_1^n, y_2^n | x_1^n, x_2^n, x_3^n) = \prod_{i=1}^{n} p(y_{1_i}, y_{2_i} | x_{1_i}, x_{2_i}, x_{3_i}) \quad (1)$$

for every discrete time instant i synchronously.

*Definition 1:* The discrete memoryless IC-DMS-UDC is described by a tuple $(\mathcal{X}_1, \mathcal{X}_2, \mathcal{X}_3, \mathcal{Y}_1, \mathcal{Y}_2, p(y_1, y_2 | x_1, x_2, x_3))$, where $\mathcal{X}_1, \mathcal{X}_2$ and $\mathcal{X}_3$ denote the channel input alphabets, $\mathcal{Y}_1$ and $\mathcal{Y}_2$ denote the channel output alphabets, and $p(y_1, y_2 | x_1, x_2, x_3)$ denotes the transition probability.

Next we present the following definitions with regard to the existence of codes, achievable rate regions and capacity region for the discrete memoryless IC-DMS-UDC.

*Definition 2:* A $(|W_1|, |W_2|, n, P_e)$ code consists of:
1. An encoding function for sender 1:
   $f_1: W_1 \to X_1^n$.
2. An encoding function for sender 2:
   $f_2: W_1 \times W_2 \to X_2^n$.
3. A family of encoding functions for receiver 2:
   $f_{3_i}: (Y_{2_1}, Y_{2_2}, Y_{2_3}, \ldots, Y_{2_{i-1}}) \to X_{3_i}$ for $1 \leq i \leq n$.
   Note that this family of encoding functions is much different from $f_1$ and $f_2$.
4. A decoding function for receiver 1:
   $g_1: Y_1^n \to \widehat{W}_1$.
5. A decoding function for receiver 2:
   $g_2: Y_2^n \to \widehat{W}_2$.
6. The average probability of error:
   $P_e := \max\{P_{e_1}, P_{e_2}\}$,
   where $P_{e_j}$ denotes the average probability of error at decoder j, j = 1,2. Furthermore, it is assumed that all the $(w_1, w_2) \in (W_1, W_2)$ are equiprobable. So $P_{e_j}$ can be computed as

$$P_{e_j} = \frac{1}{|W_1| \times |W_2|} \sum_{(w_1,w_2) \in (W_1,W_2)} \Pr\{\widehat{w}_j \neq w_j | (w_1, w_2) \text{ sent.}\} \quad (2)$$

*Definition 3:* A nonnegative rate pair $(R_1, R_2)$ is said to be achievable for the IC-DMS-UDC if there exists a sequence of $(|W_1|, |W_2|, n, P_e)$ codes with $R_1 \leq \frac{1}{n} \log |W_1|$ and $R_2 \leq \frac{1}{n} \log |W_2|$ such that $P_e \to 0$ as $n \to \infty$. Further, the capacity region for the IC-DMS-UDC, denoted as $\mathcal{C}$, is the set of all the achievable rate pairs. And an achievable region, denoted as $\mathcal{R}$, is a subset of the capacity region.

## III. AN ACHIEVABLE RATE REGION FOR THE DISCRETE MEMORYLESS IC-DMS-UDC

In this section, we present an achievable rate region for the discrete memoryless IC-DMS-UDC, the main result of this paper.

*Theorem 1:* Let $\mathcal{P}$ denote the set of all joint probability distributions $p(q, t, x_3, v, s, x_1, u, w, x_2, y_1, y_2)$ which can be factored in the following form:

$$p(q)p(t|q)p(x_3|t,q)p(v|t,q)p(s|v,t,q)p(x_1|s,v,t,q) \\ \times p(u|s,v,t,q)p(w|s,v,t,q)p(x_2|u,w,s,v,t,q) \\ \times p(y_1, y_2 | x_1, x_2, x_3) \quad (3)$$

Let $\mathcal{R}(p)$ be the set of all nonnegative rate pairs $(R_1, R_2) = (R_{11} + R_{1R}, R_{22} + R_{2P})$ such that the following inequalities hold:

$$R_{11} \leq I(S; W, Y_1 | V, T, Q) \quad (4)$$
$$R_{2P} \leq I(W; Y_1 | T, Q) - I(V, S; W | T, Q) \quad (5)$$
$$R_{11} + R_{1R} + R_{2P} \leq I(T, W, V, S; Y_1 | Q) \quad (6)$$
$$R_{1R} \leq I(V; U, W, Y_2 | T, Q) \quad (7)$$
$$R_{22} \leq I(U; V, W, Y_2 | T, Q) - I(V, S; U | T, Q) \quad (8)$$
$$R_{2P} \leq I(W; V, U, Y_2 | T, Q) - I(V, S; W | T, Q) \quad (9)$$

$$R_{1R} + R_{22} \le I(V,U;W,Y_2|T,Q) + I(U;V|T,Q) \\ - I(V,S;U|T,Q) \quad (10)$$

$$R_{1R} + R_{2P} \le I(V,W;U,Y_2|T,Q) + I(W;V|T,Q) \\ - I(V,S;W|T,Q) \quad (11)$$

$$R_{22} + R_{2P} \le I(U,W;V,Y_2|T,Q) + I(U;W|T,Q) \\ - I(V,S;U|T,Q) - I(V,S;W|T,Q) \quad (12)$$

$$R_{1R} + R_{22} + R_{2P} \le I(V,U,W;Y_2|T,Q) + I(V;U,W|T,Q) \\ + I(W;U|T,Q) - I(V,S;U|T,Q) - I(V,S;W|T,Q) \quad (13)$$

Then the region $\mathcal{R}(p)$ is an achievable rate region for the discrete memoryless IC-DMS-UDC.

*Proof:* For separately encoding, the message $W_1$ of $nR_1B$ bits is split into two parts, $W_{11}$ of $nR_{11}B$ bits and $W_{1R}$ of $nR_{1R}B$ bits. In the symmetric manner, $W_2$ of $nR_2B$ bits is split into two parts, $W_{22}$ of $nR_{22}B$ bits and $W_{2P}$ of $nR_{2P}B$ bits. And the message $W_{11}$ is split into B equally-sized sub-messages $(W_{11_1}, W_{11_2}, \ldots, W_{11_B})$ each with $nR_{11}$ bits. In exactly the same manner, $W_{1R}$, $W_{22}$ and $W_{2P}$ are split into B equally-sized sub-messages respectively. It's assumed that the transmission will be completed in $n \times (B+1)$ symbols (or equivalently, $B+1$ blocks). Encoding, transmission and decoding will be performed in $(B+1)$ blocks. Note that as $B \to \infty$, the rate pair $(R_1, R_2) = \left(\frac{B(R_{11}+R_{1R})}{B+1}, \frac{B(R_{22}+R_{2P})}{B+1}\right)$ is arbitrarily close to $(R_{11}+R_{1R}, R_{22}+R_{2P})$.

We shall construct "separate codebooks" for each block. (In the rest of this paper, we use terminal 1, 2, 3 and 4 to refer to sender 1, sender 2, receiver 1 and receiver 2 respectively.)

**Codebook Construction:**

For block b, $b = 1,2,3,\ldots,B+1$, we generate a sequence $q_b^n$ by choosing the symbols independently using $P(q)$ and show this $q_b^n$ to terminal 1, 2, 3 and 4. Note that Q is a time-sharing random variable used to strictly extend the achievable region obtained by convex hull operation [6]. And we construct two codebooks for each block b.

**Codebook 1:** We employ partial-decode-and-forward [2, Ch. 9.4] strategy to construct codebook 1. Consider two distributions $P(x_1, s, v, t|q)$ and $P(x_3|t, q)$. For $q_b^n$, generate $2^{nR_{1R}}$ $t_b^n(w_{1R_{b-1}})$, $w_{1R_{b-1}} = 1,2,\ldots,2^{nR_{1R}}$, by choosing each symbol independently using $P(t|q)$. And for each $(t_b^n(w_{1R_{b-1}}), q_b^n)$, generate one codeword $x_{3_b}^n(w_{1R_{b-1}})$ by choosing each symbol independently using $P(x_3|t, q)$.

Next, for each $(t_b^n(w_{1R_{b-1}}), q_b^n)$, use superposition coding to generate $2^{nR_{1R}}$ $v_b^n(w_{1R_{b-1}}, w_{1R_b})$, $w_{1R_b} = 1,2,\ldots,2^{nR_{1R}}$, by choosing each symbol independently using $P(v|t, q)$. And for each $(v_b^n(w_{1R_{b-1}}, w_{1R_b}), t_b^n(w_{1R_{b-1}}), q_b^n)$, use superposition coding to generate $2^{nR_{11}}$ $s_b^n(w_{1R_{b-1}}, w_{1R_b}, w_{11_b})$, $w_{11_b} = 1,2,\ldots,2^{nR_{11}}$, by choosing each symbol independently using $P(s|v, t, q)$.

Lastly for each $(s_b^n(w_{1R_{b-1}}, w_{1R_b}, w_{11_b}), v_b^n(w_{1R_{b-1}}, w_{1R_b}), t_b^n(w_{1R_{b-1}}), q_b^n)$, generate one codeword $x_{1_b}^n(w_{1R_{b-1}}, w_{1R_b}, w_{11_b})$ by choosing each symbol independently using $P(x_1|s, v, t, q)$.

**Codebook 2:** We employ Gel'fand-Pinsker coding [5] strategy to construct codebook 2. Consider two distributions $P(u|t, q)$ and $P(w|t, q)$. For each $(t_b^n(w_{1R_{b-1}}), q_b^n)$, generate $2^{n(R_{22}+R'_{22})}$ $u_b^n$ by choosing each symbol independently using $P(u|t, q)$. Then put these $2^{n(R_{22}+R'_{22})}$ sequences into $2^{nR_{22}}$ bins uniformly such that each bin contains $2^{nR'_{22}}$ sequences. And index each bin as $w_{22_b}$, $w_{22_b} \in \{1,2\ldots,2^{nR_{22}}\}$. In the symmetric manner, generate $2^{n(R_{2P}+R'_{2P})}$ $w_b^n$ by choosing each symbol independently using $P(w|t, q)$. Put these $2^{n(R_{2P}+R'_{2P})}$ sequences into $2^{nR_{2P}}$ bins uniformly such that each bin contains $2^{nR'_{2P}}$ sequences. And index each bin as $w_{2P_b}$, $w_{2P_b} \in \{1,2\ldots,2^{nR_{2P}}\}$.

**Encoding and Transmission:**

Encoding and transmission are performed in $B+1$ blocks. Note that terminal 4 serves a dual purpose: working as the receiver for sender 2 and the relay for sender 1.

**Encoding and Transmission for Terminal 1:** In block b, in order to send $(W_{11_b}, W_{1R_b}) = (w_{11_b}, w_{1R_b})$, terminal 1 transmits $x_{1_b}^n(w_{1R_{b-1}}, w_{1R_b}, w_{11_b})$. Note that for mathematical symmetry, we assign $w_{11_0} = w_{1R_0} = w_{11_{B+1}} = w_{1R_{B+1}} = 1$.

**Encoding and Transmission for Terminal 2:** In block b, in order to send $(W_{22_b}, W_{2P_b}) = (w_{22_b}, w_{2P_b})$, terminal 2 looks into bin $w_{22_b}$ to find a sequence $u_b^n$ such that

$$(t_b^n(w_{1R_{b-1}}), v_b^n(w_{1R_{b-1}}, w_{1R_b}), s_b^n(w_{1R_{b-1}}, w_{1R_b}, w_{11_b}), \\ u_b^n \text{ in bin } w_{22_b}, q_b^n) \in T_\epsilon^n(V, S, U|T, Q)$$

In the symmetric manner, terminal 2 looks in bin $w_{2P_b}$ to find a sequence $w_b^n$ such that

$$(t_b^n(w_{1R_{b-1}}), v_b^n(w_{1R_{b-1}}, w_{1R_b}), s_b^n(w_{1R_{b-1}}, w_{1R_b}, w_{11_b}), \\ w_b^n \text{ in bin } w_{2P_b}, q_b^n) \in T_\epsilon^n(V, S, W|T, Q).$$

If there is no such $u_b^n$ and $w_b^n$, declare an encoding error. Then for the two found sequences $(u_b^n, w_b^n)$ and
$(t_b^n(w_{1R_{b-1}}), v_b^n(w_{1R_{b-1}}, w_{1R_b}), s_b^n(w_{1R_{b-1}}, w_{1R_b}, w_{11_b}), q_b^n)$, generate one $x_{2_b}^n$ by choosing each symbol independently using $P(x_2|s, v, t, u, w, q)$. And transmit $x_{2_b}^n$. Note that for mathematical symmetry, we assign $w_{22_0} = w_{2P_0} = w_{22_{B+1}} = w_{2P_{B+1}} = 1$.

**Encoding and Transmission for Terminal 4:** Note that encoding, transmission and decoding processes of terminal 4 are combined together. We shall state its operation later.

**Decoding:**

Note that decoding is always performed after the transmission of block b, $b = 1,2,3,\ldots,B+1$.

**Decoding for Terminal 3:** Sliding-window decoding [2, Section 9.2] is employed at terminal 3. After the transmission of block b, terminal 3 has seen $y_{1_{b-1}}^n$ and $y_{1_b}^n$. It tries to find a tuple $(\hat{w}_{1R_{b-1}}, \hat{w}_{11_{b-1}}, \hat{w}_{2P_b})$ such that

$$(t_{b-1}^n(\tilde{w}_{1R_{b-2}}), v_{b-1}^n(\tilde{w}_{1R_{b-2}}, \hat{w}_{1R_{b-1}}), \\ s_{b-1}^n(\tilde{w}_{1R_{b-2}}, \hat{w}_{1R_{b-1}}, \hat{w}_{11_{b-1}}), \\ w_{b-1}^n \text{ in bin } \tilde{w}_{2P_{b-1}}, y_{1_{b-1}}^n, q_{b-1}^n) \in T_\epsilon^n(S, V, Y_1|W, T, Q)$$

and
$$(t_b^n(\hat{w}_{1R_{b-1}}), w_b^n \text{ in bin } \hat{w}_{2P_b}, y_{1_b}^n, q_b^n) \in T_\epsilon^n(T, W, Y_1|Q)$$
where $\tilde{w}_{1R_{b-2}}$ and $\tilde{w}_{2P_{b-1}}$ are the estimates in the previous block. If there is one or more such tuples, terminal 3 chooses one and declares $(\hat{w}_{1R_{b-1}}, \hat{w}_{11_{b-1}}, \hat{w}_{2P_b})$ was sent. Otherwise a decoding error is declared.

**Encoding, Transmission and Decoding for Terminal 4:**
After the transmission of block b, terminal 4 has seen $y_{2_b}^n$. It tries to find a tuple $(\hat{w}_{1R_b}, \hat{w}_{22_b}, \hat{w}_{2P_b})$ such that
$$(t_b^n(\tilde{w}_{1R_{b-1}}), v_b^n(\tilde{w}_{1R_{b-1}}, \hat{w}_{1R_b}), u_b^n \text{ in bin } \hat{w}_{22_b},$$
$$w_b^n \text{ in bin } \hat{w}_{2P_b}, y_{2_b}^n, q_b^n) \in T_\epsilon^n(V, U, W, Y_2|T, Q)$$
where $\tilde{w}_{1R_{b-1}}$ is the estimate in the previous block. If there is one or more such tuples, terminal 4 chooses one and declares $(\hat{w}_{1R_b}, \hat{w}_{22_b}, \hat{w}_{2P_b})$ was sent. Otherwise a decoding error is declared.

After terminal 4 estimates $\hat{w}_{1R_b}$, it transmits $x_{3_{b+1}}^n(\hat{w}_{1R_b})$ in the next block.

**Probability of Error Analysis:**
*(Presented in Appendix)* ∎

We next extend the obtained achievable rate region to two special cases.

## IV. TWO SPECIAL CASES OF THEOREM 1

From an information-theoretic perspective, IC-DMS-UDC includes IC-DMS as a special case. More specifically, by choosing $X_3$ as ∅, IC-DMS-UDC is equivalent to IC-DMS in all aspects. Hence, by choosing $X_3$ as ∅, any achievable rate regions for IC-DMS-UDC are also achievable for IC-DMS. In this section, we consider two special cases of Theorem 1. First, we derive an achievable rate region for IC-DMS from Theorem 1. Remarkably, this region is proved to be the capacity region for the Gaussian IC-DMS in the low-interference-gain regime [10], [12]. Second, the Gaussian IC-DMS-UDC is considered and a numerical example is provided to demonstrate the improvements.

### A. An Achievable Rate Region for the Discrete Memoryless IC-DMS

*Corollary 1:* Let $\mathcal{P}^*$ denote the set of all joint probability distributions $p(q, s, x_1, u, x_2, y_1, y_2)$ which can be factored in the following form:
$$p(q)(s|q)p(x_1|s, q)p(u|s, q)p(x_2|u, s, q)$$
$$\times p(y_1, y_2|x_1, x_2) \quad (14)$$
Let $\mathcal{R}^*(p)$ be the set of all nonnegative rate pairs $(R_1, R_2)$ such that the following inequalities hold:
$$R_1 \le I(S; Y_1|Q) \quad (15)$$
$$R_2 \le I(U; Y_2|Q) - I(S; U|Q) \quad (16)$$
Then the region $\mathcal{R}^*(p)$ is an achievable rate region for the discrete memoryless IC-DMS.

*Proof:* By choosing $X_3, V, W$ as ∅, this corollary can be proved accordingly. ∎

Now we consider the Gaussian IC-DMS-UDC.

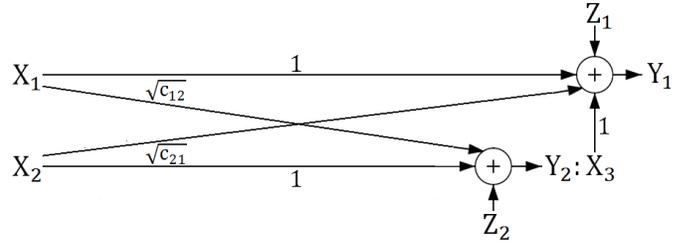

Fig. 2. The Gaussian IC-DMS-UDC.

### B. The Gaussian IC-DMS-UDC

According to [3, Ch. 7], the achievable rate regions derived in Theorem 1 and Corollary 1 can be extended to discrete (time) memoryless channel with continuous alphabets by quantizing the input, output, and interference variables. Here we focus on the Gaussian IC-DMS-UDC, depicted in Fig. 2. This channel can be described as
$$\begin{cases} Y_1 = X_1 + \sqrt{c_{21}}X_2 + X_3 + Z_1 \\ Y_2 = X_2 + \sqrt{c_{12}}X_1 + Z_2 \end{cases} \quad (17)$$
where $Z_i, i = 1,2, \sim \mathcal{N}(0,1)$, where $\mathcal{N}(0,1)$ denotes a Gaussian random variable with zero mean and unit variance, and $\sqrt{c_{12}}$ and $\sqrt{c_{21}}$ are the normalized link gains. In addition, the transmitted codewords $x_i^n, i = 1,2,3$, is subject to an average power constraint given by $E[X_i^2] \le P_i, i = 1,2,3$. First we map the random variables involved in the joint distribution (3) to a set of Gaussian random variables. Consider the case where $\alpha, \beta, \lambda_1, \lambda_2, \lambda_3 \in [0,1]$ and $\gamma_1, \gamma_2 \in [0, \infty)$ and additional auxiliary random variables $A, B, C, \tilde{U}, \tilde{W}$. And the following relations hold:

1. Set the time-sharing random variable Q as a constant.
2. The random variables $A \sim \mathcal{N}(0,1)$, $B \sim \mathcal{N}(0,1)$, $\tilde{U} \sim \mathcal{N}(0, \alpha\beta P_2)$ and $\tilde{W} \sim \mathcal{N}(0, \alpha(1-\beta)P_2)$.
3. $C = \left(\sqrt{\lambda_1}A + \sqrt{(1-\lambda_1)}B\right)$.
4. $T = \left(\sqrt{P_3}\right)A$.
5. $S = \left(\sqrt{P_1}\right)C$.
6. $V = S$.
7. $U = \tilde{U} + \sqrt{\lambda_2}T_3 + \gamma_1 T_1$.
8. $W = \tilde{W} + \sqrt{\lambda_3}T_3 + \gamma_2 T_1$.
9. $X_1 = S$.
10. $X_2 = \tilde{U} + \tilde{W} + \sqrt{(1-\alpha)P_2}C$.
11. $X_3 = T$.

With these relations above, we can obtain the covariance matrix between all random variables. Then the achievable rate region derived in Theorem 1 can be extended to the Gaussian case by calculating the mutual information terms in (4)–(13). Due to limited space, the detailed derivations are omitted.

Next we provide a numerical example to compare the Gaussian IC-DMS-UDC with the Gaussian IC-DMS. In order to have fair comparisons, we impose two additional constraints:
$$P_1 \text{ in IC-DMS} = P_1 + P_3 \text{ in IC-DMS-UDC}$$
$$P_2 \text{ in IC-DMS} = P_2 \text{ in IC-DMS-UDC}$$
Furthermore, the interference parameter $\sqrt{c_{21}}$ is intentionally chosen in the range [0,1] because the capacity region of the Gaussian IC-DMS in this case was known [10], [12]. In Fig. 3, two different parameter settings are provided. The green curve represents the capacity region of the Gaussian IC-DMS. And the red curve represents the achievable rate region of the Gaussian IC-DMS-UDC derived in this paper. As can be seen in this figure, by allowing power allocation between sender 1 and receiver 2, our achievable rate region proved in Theorem 1 offers considerable improvements under two different parameter settings.

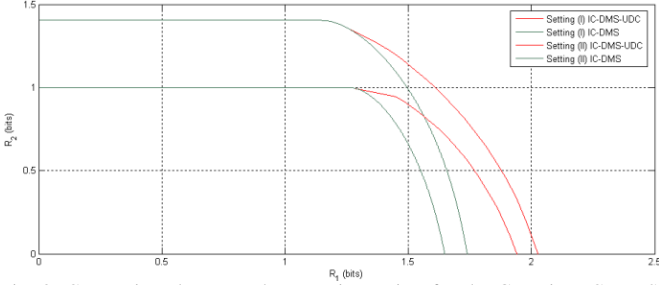

Fig. 3. Comparison between the capacity region for the Gaussian IC-DMS and our achievable rate region for the Gaussian IC-DMS-UDC in two different parameter settings: (I) $P_1 = P_2 = 6$ in the IC-DMS; $P_1 + P_3 = P_2 = 6$ in the IC-DMS-UDC, $\sqrt{c_{21}} = 0.3$, $\sqrt{c_{12}} = 2$; (II) $P_1 = 6$, $P_2 = 3$ in the IC-DMS; $P_1 + P_3 = 6$, $P_2 = 3$ in the IC-DMS-UDC, $\sqrt{c_{21}} = 0.3$, $\sqrt{c_{12}} = 2$.

## V. CONCLUSION

In this paper, a coding scheme is developed and the associated achievable rate region is derived for the IC-DMS-UDC. A Gaussian example is provided to demonstrate that by allowing power allocation between sender 1 and receiver 2, the achievable rate region can be significantly expanded.

## APPENDIX
## PROBABILITY OF ERROR ANALYSIS

Consider $P_e$, the average probability of error. Since it's assumed that all the messages $((w_{11}, w_{1R}, w_{22}, w_{2P})) \in ((W_{11}, W_{1R}, W_{22}, W_{2P}))$ are equiprobable, without loss of generality, we assume that $(m_{11}, m_{1R}, m_{22}, m_{2P})$ is sent.

Note that decoding is performed in $(B+1)$ blocks. By the same approach in [2, Section 9.2], $P_e$ can be bounded as

$$P_e \leq (B+1) \times \max_{b=1,2,\ldots,B+1} (\Pr\{\text{Error occurs in block } b \mid [\text{No error was made up to block } b.] \cap [(m_{11}, m_{1R}, m_{22}, m_{2P}) \text{ sent.}]\}) \quad (18)$$

We define several events at the bottom of this page. Besides, we use $\overline{E}$ to denote the complement of the event $E$. We bound the latter factor on the RHS in (18) as the following:

$$\Pr\{\text{Error occurs in block } b \mid E_b\} \leq \Pr\{\overline{E_{0_b}^U} \mid E_b\} + \Pr\{\overline{E_{0_b}^W} \mid E_b\}$$
$$+ \Pr\left\{\overline{E_{\text{Joint}_b}^1(m_{1R_{b-1}}, m_{11_{b-1}}, m_{2P_b})} \mid E_b \cap E_{0_b}^U \cap E_{0_b}^W\right\}$$
$$+ \sum_{(\hat{\imath} \neq m_{1R_{b-1}}, \hat{\jmath} \neq m_{11_{b-1}}, \hat{k} \neq m_{2P_b})} \Pr\{E_{\text{Joint}_b}^1(\hat{\imath}, \hat{\jmath}, \hat{k}) \mid E_{T_b}^1\}$$
$$+ \Pr\left\{\overline{E_{\text{Joint}_b}^2(m_{1R_b}, m_{22_b}, m_{2P_b})} \mid E_b \cap E_{0_b}^U \cap E_{0_b}^W\right\}$$
$$+ \sum_{(\hat{\imath} \neq m_{1R_b}, \hat{\jmath} \neq m_{22_b}, \hat{k} \neq m_{2P_b})} \Pr\{E_{\text{Joint}_b}^2(\hat{\imath}, \hat{\jmath}, \hat{k}) \mid E_{T_b}^2\} \quad (19)$$

The six terms in (19) can be bounded separately and several conditions will be evaluated by typicality arguments. After combing these conditions and letting B become infinity, Theorem 1 can be proved accordingly.

**Definitions of Several Events:**

$E_b = \{[\text{No error was made up to block } b.] \cap [(m_{11}, m_{1R}, m_{22}, m_{2P}) \text{ sent.}]\}$

$E_{0_b}^U = \{(t_b^n(m_{1R_{b-1}}), v_b^n(m_{1R_{b-1}}, m_{1R_b}), s_b^n(m_{1R_{b-1}}, m_{1R_b}, m_{11_b}), u_b^n \text{ in bin } m_{22_b}, q_b^n) \in T_\epsilon^n(V, S, U|T, Q)\}$

$E_{0_b}^W = \{(t_b^n(m_{1R_{b-1}}), v_b^n(m_{1R_{b-1}}, m_{1R_b}), s_b^n(m_{1R_{b-1}}, m_{1R_b}, m_{11_b}), w_b^n \text{ in bin } m_{2P_b}, q_b^n) \in T_\epsilon^n(V, S, W|T, Q)\}$

$E_{\text{Joint}_b}^1(\hat{w}_{1R_{b-1}}, \hat{w}_{11_{b-1}}, \hat{w}_{2P_b}) = \{[(t_{b-1}^n(m_{1R_{b-2}}), v_{b-1}^n(m_{1R_{b-2}}, \hat{w}_{1R_{b-1}}), s_{b-1}^n(m_{1R_{b-2}}, \hat{w}_{1R_{b-1}}, \hat{w}_{11_{b-1}}), w_{b-1}^n \text{ in bin } m_{2P_{b-1}}, y_{1_{b-1}}^n, q_{b-1}^n) \in T_\epsilon^n(V, S, Y_1|T, W, Q)]$
$\cap [(t_b^n(\hat{w}_{1R_{b-1}}), w_b^n \text{ in bin } \hat{w}_{2P_b}, y_{1_b}^n, q_b^n) \in T_\epsilon^n(T, W, Y_1|Q)]\}$

$E_{\text{Joint}_b}^2(\hat{w}_{1R_b}, \hat{w}_{22_b}, \hat{w}_{2P_b}) = \{(t_b^n(m_{1R_{b-1}}), v_b^n(m_{1R_{b-1}}, \hat{w}_{1R_b}), u_b^n \text{ in bin } \hat{w}_{22_b}, w_b^n \text{ in bin } \hat{w}_{2P_b}, y_{2_b}^n, q_b^n) \in T_\epsilon^n(V, U, W, Y_2|T, Q)\}$

$E_{T_b}^1 = \{E_b \cap E_{0_b}^U \cap E_{0_b}^W \cap E_{\text{Joint}_b}^1(m_{1R_{b-1}}, m_{11_{b-1}}, m_{2P_b})\}$

$E_{T_b}^2 = \{E_b \cap E_{0_b}^U \cap E_{0_b}^W \cap E_{\text{Joint}_b}^2(m_{1R_b}, m_{22_b}, m_{2P_b})\}$